\documentstyle[sprocl]{article}

\def\beq{\begin{equation}}
\def\eeq{\end{equation}}

\def\be{\begin{equation}}
\def\ee{\end{equation}}
\def\bea{\begin{eqnarray}}
\def\eea{\end{eqnarray}}

\begin{document}

\title{QCD DYNAMICS FROM M-THEORY}

\author{Nick Evans}

\address{Department of Physics, Boston University, 590 Commonwealth Ave,\\
Boston, MA 02215, USA\\E-mail: nevans@physics.bu.edu}

\maketitle\abstracts{The field theories on the surface of
  non-supersymemtric D-brane constructions are identified. By moving to
  M-theory a semi-classical, strong coupling expansion to the IR 
  non-supersymmetric gauge dynamics is obtained. The solution is
  consistent with the formation of a quark condensate but there is
  evidence that in moving to strong coupling scalar degrees of freedom
  have not decoupled.}

\section{D-Brane Constructions of Field Theories}

A large literature has grown up on engineering supersymmetric
field theories using D-brane \cite{Dnotes} constructions in type IIA string
theory (a comprehensive review and list of references is to be found in
ref 2). 
The essential ingredient is that the massless string modes of strings
ending on the surfaces of D-branes correspond to gauge fields living on
the D-branes' surfaces, plus superpartners. In addition the D-branes break
half the supersymmetries of the background space-time (the precise
generators depending upon their orientation). In this section we will
use these properties in perturbative string theory to construct field
theories in 4D with N=2, N=1 and N=0 supersymmetries.
Whilst a perturbative identification between the string states in the
type IIA theory and the UV field theory may be made, the IIA picture
provides no information about the strongly coupled 
IR dynamics of the theory. Such information would correspond to short
distance structure of the branes, but due to the strongly coupled nature
of the core of NS5 branes, the precise structure of an NS5 D4 junction is
unclear. In section 2 by moving to M-theory \cite{witten} 
a strong coupling expansion
will be made allowing some understanding of strong coupling.

The basic brane configuration from which we start corresponds
to $SU(N)$ N=2 SQCD with $F$ quark matter flavours. It is given from left to
right by the branes

\begin{equation}
\begin{tabular}{|c|c|c|c|c|c|c|c|c|}
\hline
 & $\#$ & $R^4$ & $x^4$ & $x^5$ &  $x^6$ & $x^7$ & $x^8$ & $x^9$ \\
\hline 
NS5 & 1 & $-$ & $-$ & $-$  &  $\bullet$ & $\bullet$ & $\bullet$ & $\bullet$ \\
\hline
D4 & $N$ & $-$  & $\bullet$ &  $\bullet$ &  $[-]$ & $\bullet$ & 
                              $\bullet$ & $\bullet$ \\
\hline
NS5$'$ & 1 & $-$ & $-$ & $-$ & $\bullet$ & $\bullet$  &  $\bullet$ & $\bullet$ \\
\hline
D4$'$ & $F$ & $-$  & $\bullet$ &  $\bullet$ & $-$ & $\bullet$ & 
                              $\bullet$ & $\bullet$ \\
\hline
\end{tabular} 
\end{equation}

$R^4$ is the space $x^0-x^3$. A dash $-$ represents a
direction along a brane's world wolume while a dot $\bullet$ is
transverse. For the special case of the D4-branes' $x^6$ direction,
where the world volume is a finite interval corresponding to their
suspension between two NS5 branes at different values of $x^6$, 
we use the symbol $[-]$. 
The field theory exists in the world volume of the D4 branes 
on scales much greater 
than the $L_6$ distance between the NS5 branes. The fourth
space like direction of the D4-branes generates  the coupling of the
gauge group in the  effective 4D theory. The semi-infinite D4$'$ brane
is responsible for providing the quark flavours.

The $U(1)_R$ and $SU(2)_R$ symmetries of the N=2 field theory are
manifest in the brane picture. They correspond to isometries of the
configuration; an SO(2) in the  $x^4,x^5$ directions and an SO(3) in the
$x^7, x^8, x^9$ directions.

N=2 supersymmetry may be broken in the configuration by rotations of the
branes away from this configuration of maximal symmetry
\cite{barbon1,softb,softb1}.
The positions of the branes in these configurations break supersymmetry
and hence we expect there to be supersymmetry breaking parameters introduced
in the low energy field theory lagrangian. These parameters must be the
supersymmetry breaking vevs of fields in the string theory since at tree
level there are no supersymmetry breaking parameters. The vevs occur as
parameters because the fluctuations of those fields are being neglected
in the field theory; such fields are spurions. They have a natural
interpretation in the brane configuration. The fields are those
describing the positions of the branes and their fluctuations are
neglected because the infinite branes are very massive. If though we
choose to include these fields in the field theory description they
occur subject to the stringent constraints of N=2
supersymmetry \cite{N=2spurion}. 
The
spurions whose vevs correspond to the supersymmetry breaking parameters
must be the auxilliary fields of N=2 multiplets. This constraint is
sufficient to identify the spurions \cite{softb1}.

\begin{itemize}
{\item 
The NS5 branes may be rotated
in the  $x^4,x^5,x^7, x^8, x^9$ space (rotations from the N=2
configuration into the $x^6$
direction cause the NS5 branes to cross, changing the topology of the
configuration in such a way that it can no longer be easily identified
with a field theory). These rotations correspond to components of the 
spurion fields occuring as  vector fields 
in the prepotential of the N=2 theory as ${\cal F} = (S_1 + i
S_2)A^2$ \cite{N=2spurion}. This is the unique way in which a spurion
may be introduced compatible with the N=2 supersymmetry.
The scalar spurion vevs generate the gauge coupling $\tau$.
When we allow the auxilliary fields of the spurions to be
non-zero we obtain the tree level masses

\begin{eqnarray} \label{softUV}
&& \nonumber
-{N_c \over 8  \pi^2} Im \left( (F_1^*+iF_2^*) \psi_A^\alpha \psi_A^\alpha 
+ (F_1 + i F_2)
\lambda^\alpha \lambda^\alpha + i \sqrt{2}
(D_1 + i D_2) \psi_A^\alpha\lambda^\alpha \right) \\
&&
  - {N_c \over 4 \pi^2 Im (s_1+is_2)} \left((|F_1|^2 + D_1^2/2) Im(
a^\alpha)^2
+  (|F_2|^2 + D_2^2/2) Re
(a^\alpha)^2 \right. \\
&& \left. \right. \hspace{4cm} + \left. 
 (F_1 F_2^* + F_1^*F_2 + D_1D_2) Im(a^\alpha) Re(a^\alpha)
\right)\nonumber
\end{eqnarray}
A number of consistency checks support the identification \cite{softb1}. 
Switching on
any one of the six  independent real supersymmetry breakings in the field
theory leaves the same massless spectrum in the field theory as in the
brane picture when any one of the six independent rotations of the NS5
brane is performed. The field theory and brane configurations possess
the same sub-manifold of N=1 supersymmetric configurations.}
{\item Supersymmetry may also be broken by forcing the D4 
and D4$'$ branes to lie at
angles to each other \cite{softb2}.
With the introduction of matter fields in the field theory a single
extra spurion field is introduced associated with the quark mass. The
only possibility is to promote the mass to an N=2 vector multiplet
associated with $U(1)_B$ \cite{N=2spurion}.  Switching on its
auxilliary field vevs induce the tree level supersymmetry breaking 
operators
\beq \label{scalarmass}
2 Re( F_{M}  q \tilde{q}) + D_{M} \left( |q|^2 -
|\tilde{q}|^2 \right)
\eeq
Again a number of consistency checks support the identification of these field
theory breakings with the angles between D4 branes \cite{softb2}. There are
three independent real parameters in both the field theory and the brane
picture. The scalar masses in the field theory 
break $SU(2)_R$ but leave two  $U(1)_R$
symmetries of the supersymmetric theory intact. The scalar masses may 
always be
brought to diagonal form by an $SU(2)_R$ transformation that mixes $q$
and $\tilde{q}^*$. In the resulting basis there is an unbroken $U(1)$
subgroup of $SU(2)_R$.
In the brane picture the
D4$'$ branes lie at an angle in the $x^6-x^9$ directions breaking the
$SU(2)_R$ symmetry but leaving two $U(1)_R$ symmetries unbroken. }
\end{itemize}

The NS5 branes may be rotated in such a way as to preserve $N=1$
supersymmetry \cite{barbon1} 
(corresponding to, for example, setting $F_2 = i F_1$ in
the field theory) and the resulting configuration with the adjoint
matter field decoupled is given by

\beq 
\begin{tabular}{|c|c|c|c|c|c|c|c|c|}
\hline
 & $\#$ & $R^4$ & $x^4$ & $x^5$ &  $x^6$ & $x^7$ & $x^8$ & $x^9$ \\
\hline 
NS5 & 1 & $-$ & $-$ & $-$  &  $\bullet$ & $\bullet$ & $\bullet$ & $\bullet$ \\
\hline
D4 & $N$ & $-$  & $\bullet$ &  $\bullet$ &  $[-]$ & $\bullet$ & 
                              $\bullet$ & $\bullet$ \\
\hline
NS5' & 1 & $-$ & $\bullet$ & $\bullet$ & $\bullet$ & $-$  &  $-$ & $\bullet$ \\
\hline
D4' & $F$ & $-$  & $\bullet$ &  $\bullet$ & $-$ & $\bullet$ & 
                              $\bullet$ & $\bullet$ \\
\hline
\end{tabular} 
\eeq

Finally we note that for generic angles of the NS5 branes supersymmetry
is completely broken and the gaugino is massive. Although the scalar
masses we can generate at tree level by deformations of the brane
construction are always unbounded and force the
theory to a higgs branch, in the full theory we expect radiative masses
for the scalars of order the supersymmetry breaking scale. Thus we
expect that if a large gaugino mass is introduced the theory in the IR
is non-supersymmetric QCD. Note that the construction does not
constitute a phenomenological model of QCD because the gravitons of the
theory (which we can make arbitrarily weakly coupled relative to the
gauge dynamics) still live in a 10D space-time. 
These constructions are therefore only
mathematical tools for studying gauge theories. 

\section{Strong Coupling From M-Theory}

To attempt to 
understand the strong coupling behaviour of the field theories
constructed above  we  move to M-theory \cite{witten}. 
Type IIA string theory
with coupling $g_s$ is 
11 dimensional M-theory compactified on a
circle of radius, $R \sim g_s$, which in the IR is described by weakly
coupled 11D SUGRA. In M-theory NS5 branes
and D4 branes become aspects of a single M5 brane wrapped in places on
the compact dimension. The junctions between these objects may thus be
smoothly described by a minimal area 
embedding of the M5 brane. Increasing the M-theory compactification
radius from zero allows the study of the string theory with increased
coupling at the string scale. It is therefore possible to make a strong
coupling expansion to the field theory. That is to smoothly deform the
field theory of interest to a theory with the same global symmetries and
parameters but which is fundamentally a theory of strongly interacting
strings in the $R \rightarrow \infty$ limit. For intermediate $R$ the
theory has Kaluza Klein states in addition to those of the field theory. 
We hope that by making this transition between smoothly related 
theories there is no phase transition and that the
two theories ly in the same universality class. This technique has been
used to derive the existence of gaugino condensation in N=1 super Yang
Mills theory \cite{witten} and Seiberg's duality \cite{seiberg}
\cite{branedual,schmaltz}  
in the theories with flavour amongst other
results \cite{review}. 
For these supersymmetric theories the holomorphic and BPS
properties of states in the theory contribute to making the motion to
strong coupling smooth. For non-supersymmetric theories we have no such
protection but we can hope to look for a consistent picture free of
transitions.

We must therefore minimal area embed an M5 brane in the background 11D
space with metric
\beq
 ds^2 = \sum_{i,j=0}^9 \eta_{ij} dx^i dx^j + R^2 (dx^{10})^2
\eeq
3+1 dimensions of the M5 brane are flat (the space the field theory
lives in) and we must concentrate on embedding the remaining two
dimensions of its surface in the space $\vec{X} = x^4-x^{10}$. Classically a
minimal area embedding corresponds to when the energy momentum tensor of
the two dimensional theory on the surface of the brane vanishes
\beq \label{cond}
T_{zz} = g_{ij} \partial_z X^i \partial_z X^j = 0
\eeq
We proceed by guessing solutions with the topology that we require. A
solution is \cite{softb2} 
(with $v=x^4+ix^5$, $w=x^7+ix^8$ and $t = exp(x^6 + i
x^{10}/R)$ )

\beq 
\begin{array}{c}
v = z + {\eta \over z} + {\bar{\epsilon} \over \bar{z}}, \hspace{1cm}
w = {\xi \over z} + {\bar{\lambda} \over \bar{z}}, \hspace{1cm}
t = \kappa z^{N}/(z-m)^{F}, \\
\\
x^9 = 4 \epsilon^{1/2} Re \ln z 
\end{array}
\eeq
subject to the constraint
\beq
\eta \epsilon + \xi \lambda = 0
\eeq

The curve has two $U(1)$ symmetries associated with rotations in the $v$
and $w$ planes which are broken by the parameters of the curve. The
symmetries may be restored by assigning the parameters spurious charges
\beq
\begin{array}{c|cccccccccc}
& v & w & t&  z& m& \eta & \epsilon & \xi & \lambda & \kappa\\
\hline
U(1)_v & 2 & 0 & 0 & 2 & 2 & 4 & 0 & 2 & 2 & 2(F-N)\\
U(1)_w & 0 & 2 & 0 & 0 & 0 & 0 & 0 & 2 & -2 & 0 \\
\end{array}
\eeq

These symmetries are just the $U(1)_R$ symmetries of the field theory
and may be used to identify the parameters with field theory parameters
in the limit $R \rightarrow 0$.

Thus for example we find the curve describing an N=1 $SU(N)$ field
theory with $F$ quark flavours \cite{schmaltz,N=1brane}
\beq \label{N=1curve}
v = z, \hspace{1cm} w = {  \Lambda^{b_0/N} m^{F/N} \over z}, 
\hspace{1cm} t = z^{N}m^{F-N}/(z-m)^{F} 
\eeq
Viewing the curve asymptotically and as $\Lambda
\rightarrow 0$
\begin{eqnarray}
z \rightarrow \infty \hspace{0.3cm} & w=0 \hspace{0.3cm}
          & t = v^{N-F} m^{F-N} \nonumber\\
z \rightarrow 0 \hspace{0.3cm} & v = 0 \hspace{0.3cm} 
          & t = \left( {1 \over w}\right)^N \Lambda^{b_0} m^{F-N}
\end{eqnarray}
The $U(1)_v$ and $U(1)_w$ symmetries (allowing $m$ to transform
spuriously but not $\Lambda$) are broken to 
$Z_N$ and $Z_{N-F}$ discrete subgroups as is the case for the 
$U(1)_R$ symmetries of the
field theory. The N=1 theory behaves 
like supersymmetric Yang Mills theory below the
matter field mass scale and dynamically generates a gaugino condensate.
The theory  has $N$ degenerate vacua associated with the
spontaneous breaking of the low energy $Z_{N}$ symmetry. In the curve this
corresponds to the $N$ curves in which $\Lambda^{b_0}_n = 
\Lambda^{b_0}_0 exp(2\pi in)$. In the UV these curves can be made equivalent
by a $Z_{N}$ transformation.

We may now switch on the parameter $\epsilon$ and break
supersymmetry \cite{softb2}. The curve becomes
\beq \begin{array}{c}
v = z +{\bar{\epsilon}\over \bar{z}} 
\hspace{1cm} w = {\Lambda^{b_0/N} m^{F/N}\over z}, \hspace{1cm} t =
z^{N}m^{F-N}/(z-m)^{F}, \\ \\
 x^9 = 4 \epsilon^{1/2} Re \ln z
\end{array}
\eeq
In the $R\rightarrow 0$ limit the
D4 branes lie in the $x^6$ and $x^9$ directions and the NS5
brane lying in the $w$ direction has been rotated in a non-supersymmetric
fashion into the $v$ direction.
We generically expect the field theory to have the 
supersymmetry breaking terms
\beq   
D \left( |q|^2 - |\tilde{q}|^2 \right) + m_\lambda \lambda \lambda
\eeq
 We may identify the
parameter $\epsilon$ with field theory parameters from its symmetry
charges. Requiring that the brane configuration retains, asymptotically,
the $Z_F$
symmetry remnant of $U(1)_A$, which is left unbroken by these mass terms,
forces
\beq
\epsilon =  \left( m_{\lambda}^N \Lambda^{b_0} m_Q^F \right) ^{1/N}
\eeq

Including the gaugino mass breaks the $Z_N$ symmetry of the curve and
the N vacua of the SQCD theory are no longer equivalent. This should be
compared with the field theory where, when the SQCD theory is perturbed
by a small gaugino mass, the dynamical superpotential $\Lambda^3$ implies
a potential of the form \cite{N=1spurion}
\beq
V = \Lambda^3|_F \simeq \Lambda^3 m_\lambda \cos \left[ \theta_{vac}
\right]
\eeq
where $\theta_{vac}$ is the phase of the SQCD vacuum. The potential
splits the degeneracy between the $N$ vacua. The brane picture
is consistent with that behaviour.

It is possible to take the decoupling limit for the gaugino mass by
taking $\epsilon \rightarrow \infty$. $m$ is the only R-charged parameter
remaining and there is therefore nothing that can play the role of
either a gaugino mass or condensate fitting the assumption that the
gaugino has been decoupled.
We must define a new strong scale parameter below the gaugino mass
$\Sigma = m_\lambda^{N/F} \Lambda^{b_0/F}$. 
The decoupled curve is
\beq \label{curve}
v= z + { (\bar{m} \bar{\Sigma})^{F/N}
\over \bar{z}}, \hspace{0.5cm} 
t = z^{N}m^{F-N}/(z-m)^{F},
\hspace{0.5cm} x^{9} = 4 (m \Sigma)^{F/2N} Re \ln z
\eeq
The curve has a $Z_F$ symmetry that is broken by $m$ and $\Sigma$ that
transform spuriously with charges $+2$ and $-2$ respectively. 
That is $m$ has the
charge of a quark mass parameter and $\Sigma$ the charge of a quark
condensate. This suggests the nice interpretation that the gaugino has
decoupled and the scalars radiatively with it, leaving a theory that is
non-supersymmetric QCD in the IR with a quark condensate. 

In fact the interpretation is not so clear cut \cite{barbon2,softb3}. 
To expose a problem with
the strong coupling expansion let us investigate the fate of Seiberg's
N=1 SQCD duality when supersymmetry is broken. The duality is exhibited
by the N=1 curve \cite{schmaltz}. If we make the transformations 
\begin{eqnarray}
m & \equiv & M^{N/F-N} \Lambda^{-b_0/F-N} \nonumber \\
\Lambda^{b_0} & \equiv & \tilde{\Lambda}^{-\tilde{b}_0} \nonumber \\
z & \equiv & {M^{F/F-N} \Lambda^{-b_0/F-N} \over z'}
\end{eqnarray}
the curve (\ref{N=1curve}) is transformed to
\beq
v = {M^{F/\tilde{N}} \tilde{\Lambda}^{\tilde{b}_0/\tilde{N}} \over z'},
\hspace{1cm}
w = z', \hspace{1cm}
t = { M^{F-\tilde{N}} z^{'\tilde{N}} \over (M-z')^F}
\eeq
where $\tilde{N} = F-N$ and $\tilde{b}_0 = (3\tilde{N}-F)$.

In the limit $\tilde{\Lambda} \rightarrow 0$ at fixed $M$ this curve
degenerates to the IIA configuration describing a $SU(F-N)$ gauge theory
with $F$ quark flavours of mass $M$. This theory is 
Seiberg's dual SQCD theory. Note that in terms of the electric curve
this configuration is obtained when $\Lambda \rightarrow \infty$ with
$m$ scaled to zero appropriately to keep $M$ fixed. The duality of the
field theory is a strong-weak duality.

In fact this dual nature of the curve persists with supersymmetry
breaking. For example, in the decoupled limit the transformations
\begin{eqnarray}
m & \equiv & M^{N/F-N} \Sigma^{-F/F-N} \nonumber\\
\Sigma & \equiv & \tilde{\Sigma}^{-1}\\
z' & \equiv & {m^{F/F-N} \Sigma^{-F/F-N} \over z} \nonumber
\end{eqnarray}
give the dual curve
\beq \begin{array}{c}
v= \bar{z}' + { (M \tilde{\Sigma})^{F/\tilde{N}}
\over \bar{z'}}, \hspace{1cm} 
t = z'^{\tilde{N}}M^{F-\tilde{N}}/(M-z')^{F}, \\ \\
x^{9} = -4 (M \tilde{\Sigma})^{F/2\tilde{N}} Re \ln z' 
\end{array}
\eeq

Again the dual picture emerges from the M-theory curve in the limit
where $\Sigma \rightarrow \infty$ and $m \rightarrow 0$. Surprisingly
the M-theory seems to be telling us that for massless quarks there
is a duality symmetry between an $SU(N)$ gauge theory with $F$ quarks
and an $SU(F-N)$ gauge theory with $F$ quarks. Presumably some components
of the dual meson of SQCD also survive though we have been
cavalier in the above discussion as to the boundary conditions at
infinity of the semi-infinite D4s.

The theory on the branes surface is clearly not QCD. QCD can not have a
flavour dependent duality such as the curve suggests because the theory
has no higgs branch with which to correctly reduce the gauge group as
dual quark flavours are made massive. What has gone wrong? We suggest
two possibilities. One
is that as the strong coupling scale is taken to infinity the radiative
masses of the scalars will only be of order $\Lambda$ and they can not
be considered decoupled (in fact the dual quarks are weakly coupled so
their radiative masses may be small). The other possible problem is that
the curve for some value of $m_\lambda$ has stopped being the true
vacuum of the theory and become a local minimum at a first order phase
transition. Some other curve would then  describe the true QCD
physics. 

It is amusing to speculate on whether the curve's duality in the limit
where the gaugino is decoupled can be interpreted as a duality of
non-supersymmetric theories with light (fine tuned) scalars. In fact the
anomaly matching conditions and flows with the introduction of quark
mass terms continue to hold in the field theory and its dual without a
gaugino. Duality may therefore be the result of light scalars rather
than supersymmetry (although supersymmetry is the only natural way to
stabilize massless scalars against the hierarchy problem).

The failure of the scalar fields to decouple even in the absence of
supersymmetry is awkward for the discussion of true QCD
dynamics. Ideally we would like to decouple the scalars at tree level
and not rely on radiative effects. Unfortunately the deformations of the
brane configuration do not appear to correspond to a suitable scalar mass
term. The best we could hope to achieve is to switch on masses of the
form of (\ref{scalarmass}) 
but these masses are always unbounded and trigger a higgs
branch of the theory. We do not know of anyway to give all scalars a
positive tree level mass and so their decoupling remains frustratingly
elusive.


\section*{References}
\begin{thebibliography}{99}
\bibitem{Dnotes}
      J. Polchinsky, ``TASI Lectures On D-branes'', hep-th/9611050.
\bibitem{review} 
      A. Giveon and D. Kutasov, hep-th/9802067. 
\bibitem{witten}
      E. Witten, {\it Nucl. Phys.} B {\bf 500} 3 (1997);
      E. Witten, {\it Nucl. Phys.} B {\bf 507} 658 (1997).
\bibitem{barbon1}
      J.L.F. Barbon, {\it Phys. Lett.} B {\bf 402} 59 (1997).
\bibitem{softb} 
      A. Brandhuber, J. Sonnenschein, S. Theisen and S. Yankielowicz,
      {\it Nucl. Phys.} B {\bf 502} 125 (1997); 
      A. Hanany, M.J. Strassler and A. Zaffaroni, {\it Nucl.Phys.}
      B {\bf 513} 87 (1998);
      J.L.F. Barbon and A. Pasquinucci,  hep-th/9711030.
\bibitem{softb1}
      N. Evans and M. Schwetz, hep-th/9708122.
\bibitem{softb2}
      N. Evans, hep-th/9801159.
\bibitem{N=2spurion}
      L.~\'{A}lvarez-Gaum\'{e}, J.~Distler, C.~Kounnas and 
      M.~Mari\~{n}o,   
      {\it Int. J. Mod. Phys.} A {\bf 11} 4745 (1996);
      L.~\'{A}lvarez-Gaum\'{e} and
      M.~Mari\~{n}o, {\it Int. J. Mod. Phys.} A {\bf 12} 975 (1997);
      L.~\'{A}lvarez-Gaum\'{e},  M.~Mari\~{n}o and F. Zamora, hep-th/970307; 
      L.~\'{A}lvarez-Gaum\'{e},  M.~Mari\~{n}o and F. Zamora, 
      hep-th/9707017; M. Marino and F. Zamora, hep-th/9804038;
      N.~Evans, S.D.H.~Hsu and 
      M.~Schwetz, {\it Nucl. Phys.} B {\bf 484} 124 (1997).
\bibitem{seiberg} 
      N. Seiberg, {\it Nucl. Phys.} B {\bf 435} 129 (1995).
\bibitem{branedual} 
      S. Elitzur, A. Giveon and D. Kutasov, {\it Phys. Lett.} B
      {\bf 400} 269 (1997)
\bibitem{schmaltz} 
      M. Schmaltz and R. Sundrum, hep-th/9708015.
\bibitem{N=1brane} 
      A. Brandhuber, J. Sonnenschein, S. Theisen and S. Yankielowicz,
      {\it Phys. Lett.} B {\bf 410} 27 (1997).
\bibitem{N=1spurion}
      A. Masiero and G. Veneziano, {\it Nucl. Phys.} B {\bf 249} 593 (1985);
      N.~Evans, S.D.H.~Hsu and M.~Schwetz,
      {\it Phys. Lett.} B {\bf 404} 77 (1997).
\bibitem{barbon2} 
       J.L.F. Barbon and A. Pasquinucci, hep-th/9804029.
\bibitem{softb3}
      N. Evans, hep-th/9804097.
\end{thebibliography}
\end{document}